\documentclass[12pt,a4paper]{article}
\usepackage{graphicx} 
\usepackage{bm}
\usepackage{latexsym}
\usepackage{setspace}
\usepackage{url}
\usepackage{amsfonts} 
\usepackage{fullpage}  
\usepackage{subcaption}
\usepackage[a4paper]{geometry}
\usepackage[intlimits]{amsmath}
\usepackage{float}
\RequirePackage{amsopn}
\RequirePackage{amsfonts}
\RequirePackage{amsthm}

\title{\textbf{Kant and Quantum Gravity}}
\author{\"{O}zge Ekin G\"{u}n \\ Institute of Philosophy, Freie University Berlin\\}

\date{\today}

\begin{document}
	
\maketitle	

	 In quantum gravity space and time
	 lose their status as fundamental parts of the physical reality. However, according to  Kant, space and time are the\textit{ a priori} conditions of our experience. Does Kantian characterization of these notions give constraints to quantum gravity, or does quantum gravity make Kantian characterization of space and time an invalid approach? This paper provides answers to these questions with a philosophical approach to quantum gravity.


	 \section{Introduction}
	 Quantum gravity is  changing everything we know about space and time  by defying intuitive notions we have about them. It has the potential of becoming one of the strongest theories defining the physical relations, matter and cosmology by combining quantum mechanics and general relativity.
	 
	 In this paper I present the most favorable approach to the quantum gravity,  \textit{loop quantum gravity}, and compare the notions  emerging from this approach to Kantian characterization of space and time.  In the first part of the paper, I provide a quick historical survey on  approaches to space and time which leads to  Kant's synthesis.  In the second part, I present recent  works on quantum gravity that address philosophical issues. In the final part, after pointing out the remarks on the compatibility of Kantian and quantum gravitational notions of space and time by Rovelli \cite{Rovelli16}, Butterfield and Isham \cite{Butterfield:1999ah}, I argue that the space and time emerging from quantum gravity do not give constraints to Kantian characterization of them.  Finally, I  talk about the question raised by Rovelli on the limitations of our intuitions in understanding quantum gravity. (cf. \cite{Rovelli16}, p. 120)

	 \section{Kantian Characterization of Space and Time} 
	 Kant's notion of space is a well studied synthesis of previous theories before him. Namely, he synthesizes relational and absolute theories of space and time, which go back to Plato and Aristotle. In the following I first present a brief historical evolution of these concepts and then explain Kant's notion of space and time  as they are revealed in \textit{the Critique} \cite{Kant98}. 
	 \subsection{Kant's Historical Synthesis}
	 In Timaeus    (\cite{Plato2000}, 48e-53c)  Plato argues that space is neither  a form nor a matter, since \textit{form} is ``from which it is becoming'' and \textit{matter} is ``from which it is constituted''. In other words, space is neither ``from which it is becoming'' nor it constitutes another entity as a matter. Therefore, it is a receptacle for entities  to come into being.  Aristotle defines space as ``a limit of the containing body at which it is in contact with the contained'' (\cite{Aristo83}, 4.212a, Book IV Delta ). Space or ``place''  is different from form, matter or an empty interval between  the matter.  We will see similar notions emerging in Kantian characterization of space using  Plato's notion of receptacle  and  Aristotle's  relational view.  We also come across the relational view of space in Leibniz's characterization \cite{Leibniz56},  opposing the Newtonian notion of an absolute space and time \cite{Newton66}. Leibniz defines space as something merely relative (as time is). He asserts that space is not a substance and gives a proof of it with the need for sufficient reason whereby defending the applicability of mathematical reasoning to physical and metaphysical subjects. He concludes that ``space is nothing but an order or set of relations among bodies, so that in absence of bodies space is nothing at all except the possibility of placing them''  (\cite{Leibniz56}, pp.9-10) Newton on the other hand, defends that the space is a physical reality and absolute. He argues that motion is the only medium which space can be explored. His argument is as follows: There is a real force causing real motion, which is the rotational  absolute motion. If there is an absolute motion this requires the existence of absolute space. (\cite{Newton66}, pp.11-12) This dichotomy is quite clear in the way Kant summarizes them as follows:\footnote{Referred by (\cite{Friedman94}, p. 4): Leibniz and later Wolff (Elementa Matheseos \cite{Wolff99}), as a follower and allocator of Leibniz's thought system as opposed to Newton's, described a system as Kant characterizes it in (AKK. 475.22-476.2  \cite{Kant1902a} ). Also note that the interpretation on Leibniz denying infinite divisibility of space that Kant reveals in \textit{Physical Monadology (1756)} is not a commonly accepted view.  Most of the Leibnizian philosophers in the 18th century considered space to be infinitely divisible.} 
	 
	 \begin{table}[H] 
	 	\begin{tabular}{l l}
	 		(1) \textit{ Leibnizian-Wolffian System}  & (2) \textit {Newtonian System}\\
	 		
	 		Denies that space is infinitely divisible & Asserts that space is infinitely divisible\\
	 		
	 		Denies the void space & Void space is necessary for free motion\\
	 		
	 		Attraction or universal gravitation  & Attraction or universal gravitation \\
	 		
	 		[or geometry or mechanics] is & is explained by the forces are in bodies\\
	 		
	 		imaginary things.  &  that are active at a distance and at rest \\
	 	\end{tabular}
	 \end{table}

	 Kant was a close follower of the lively discussions between Newton and Leibniz. He analyzes the opponent thought systems of the two and gives an exceptional synthesis of these two systems. 
	 
	 The most distinguished theories of space, with the discussion on Euclid's fifth postulate, were surfacing at the time Kant was writing the second edition of his \textit{Critique}. However, most mathematicians and philosophers were not aware that these theories were about to change what they know about space and at most they regarded them merely consistent geometrical theories. Moreover, even before mathematical discussions, Thomas Reid  (1764), a Scottish philosopher, offered a spherical characterization of space using common sense. We now have not only Euclidean or spherical definition of space but many spaces with all kind of curvatures that can be defined by setting the angle in  Euclid's first postulate (e.g. Riemannian Space) in addition to the spacetime, a mathematical model combining space and time into a single continuum (see below for some representations).

	 \begin{figure} [H]
	 	\centering
	 	\includegraphics[width=70mm]{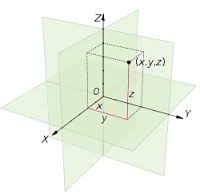}
	 	\caption{A representation for Euclidean Space \cite{Space3}}
	 \end{figure}

	 \bigskip

	 \begin{figure} [H]
	 	
	 	\includegraphics[width=150mm]{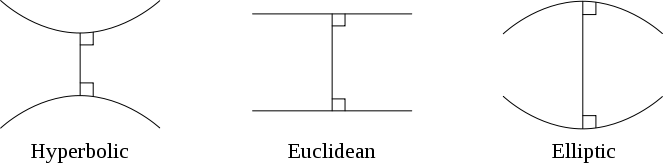} 
	 	\caption{Euclidean vs. Non-Euclidean Geometry Representations \cite{Space6}}
	 \end{figure}
	 
	 \bigskip
	 
	 \begin{figure} [H]
	 	\centering
	 	\includegraphics[width=70mm]{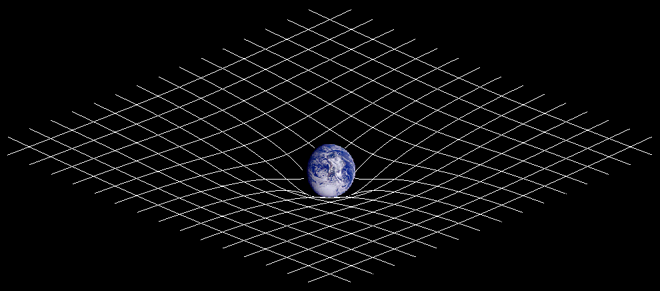} 
	 	\caption{A representation of Spacetime \cite{Space4}}
	 \end{figure}
	 
	 \bigskip
	 
	 Kant also did not consider the possibility of Euclidean space being only one of the options among many spaces. He accepted the Newtonian characterization of gravity and attraction between masses, however, thought that Newton's philosophy of nature did not reach to its limits. (cf.\cite{Friedman94}, p.1) Kant's aim then became to give a better metaphysical foundation to Newton's natural philosophy by improving it with a Leibnizian thought system. These thought systems handled  geometry and metaphysics separately, whereas Kant, from the beginning of his philosophical career set himself the task to unite the geometry and transcendental philosophy. In the following we will see how he gives a synthesis of these notions.

	 \subsection{Kantian Space and Time}
	 
	 In late 18th century Immanuel Kant was articulating the approach that shapes have to conform to the axioms and rules of Euclidean space. According to Kant, space and time are pure intuitions, meaning that they  are the results of the formation of  human mind and is acquired independently from experience.(cf.\cite{Kant98}, A26/ B42). Mathematical constructions, which are \textit{a priori} intuitions, are possible due to pure intuitions space and time.   Space and time are  forms of sensibility and  \textit{a priori} conditions of one's experience, and this space is Euclidean. For this reason, all figures that are exhibited in this space should conform to the Euclidean rules.

	 Let us deconstruct the above notions to understand what Kant means by `intuition', `\textit{a priori}' and `pure'. 
	 
	 In Kantian thought system a representation is either a concept or an intuition. When a representation is general and mediate, meaning that it is inferred from other concepts or intuitions, it is a \textit{concept}. When a representation is singular and immediate, meaning that it is not inferred from other concepts or intuitions, it is an \textit{intuition}. According to Kant, one cannot know what \textit{is} an object, she can only have an immediate or mediate representation of it. \textit{A priori} \textit{intuition} is a particular representation of a general concept where necessary properties from the concept are brought together for the purpose of construction, independent from experience.  This is neither an empirical intuition, nor a concept that has each and all the properties of, say, for example, a  triangle. What makes this \textit{a priori} is the exhibition of it, what makes it intuition is the singularity of it. Whereas \textit{a priori} intuition is exhibited independent from experience, pure intuition is \textit{acquired} without borrowing anything from experience. (\cite{Kant98}, A26/B42)

	 Accordingly, all the empirical intuitions are possible because the space is the form of sensibility and  the underlying pure intuition where all the intuitions are exhibited. It plays the role of ``basis'' for the intuitions, as the Euclidean space plays the role of ``basis'' for the figures to be exhibited in geometry. Moreover, this ``prior to all appearances space'' contains necessarily the  geometrical principles and their relations. It is the subjective condition for one's experience and perception. However, by exposing it as prior to all experience and intuitions, it becomes an objective, universal rule carrying ground. Since geometry is assumed to be an established solid ground in 18th century, and Euclidean space is the only known form of real space, by asserting the Euclidean space, as the space which already constituted by the rules of Euclidean geometry, Kant claims that he objectively reveals space as the pure basis of his transcendental philosophy.

	 Kant points out that time is also pure intuition and one can think about it without needing any appearance but it ``is not possible to think about appearances that are not in time'' (\cite{Kant98}, B46).\footnote{We will see that quantum gravity provides a counter example to this claim.} He argues that time cannot be an empirical concept and that it cannot be “drawn from experience. For, we could not experience events as simultaneous or as one-after-another unless we had an underlying a priori representation of time.” (\cite{Kant98}, B46) Hence, it must be prior to any experience and concept. Time itself is neither an object of empirical intuition nor a concept. Kant's theory also denies an absolute and trancendentally real, out- of- epistemic-reach, concept of time.(cf. \cite{Kant98}, B54)

	 After discussing the quantum gravity and the status of space and time emerging from the unification of quantum mechanics and general relativity, I will discuss how the descriptions of these notions affects Kantian characterization.  We will see that there might be some parts of his theory  that can be preserved. But first, let us try to understand how quantum gravity is reshaping everything we know about space and time.

	 \section{Quantum Gravity}
	 
	 Physicists and philosophers of physics are quite excited about the subject of quantum  gravity. This is because, in the first place, quantum gravity has the potential to unify general relativity and quantum mechanics into a consistent theory and, secondly,  the subject of quantum gravity is  incredibly rich  with philosophical issues ranging from  having  quantized gravity being consistent with black hole singularity hypothesis,   providing a quantum cosmological view of the universe where everything is made up of quanta,   to   the emerging nature of space and time. (\cite{Butterfield:1999ah}, \cite{Butterfield:1998dd}, \cite{Martinetti:2012jh}, \cite{Rovelli16}, \cite{Wuthrich2012})

	 First and foremost, the subject of quantum gravity attempts to describe gravity in terms of quantum mechanics which means that it tackles with the issue of explaining the effects of gravity in very small volume and very high energy situations which is the interpreted state of the universe right after Big Bang where the energy was  so high in magnitude compared to the almost zero volume. Planck Scale describes the region where quantum effects of gravity are expected to dominate which occurs, approximately,  at length $10^{-35}$m (compare it to, for example, the diameter of an atom, which is $10^{-10}$m), at energy $10^{22}$ GeV  and at time $10^{-42}$ seconds. (cf.{\cite{Butterfield:1999ah}, p.7) Moreover, quantum gravity is still not a well defined theory. There is not a unified approach to its definition but ranges of approaches.  However, it is a very promising area because of the unification potential I mentioned earlier. As in every breakthrough in physics and every fruitful construction in philosophy, when it seems like there is a chance of two brilliant theories to be merged, the research flares and as Kant did with the notions of space and time by combining the approaches of Newton and Leibniz, and as Newton did by combining theories of Kepler and Galileo, the new field becomes very promising and full of potential. 
	
	In the rest of this section, I describe the quantum loop gravity, currently the most favourable approach to quantum gravity.
	
	\subsection{Loop Quantum Gravity}

	To some ``loop quantum gravity is an attempt to define a quantization of gravity paying special attention to the conceptual lessons of general relativity'' (\cite{Perez:2012wv}, p. 7), to others it does  not have to be about the quantization of gravity but should be ``at least conceivable that such a theory marries a classical understanding of gravity  with a  quantum understanding of matter'' (\cite{Wuthrich2012}, pp.1-2) In the following I attempt to describe loop quantum gravity without delving into technical details and leaving aside the mathematical formulation. I combine the approaches of Rovelli \cite{Rovelli04}, \cite{Rovelli16}, Wuthrich \cite{Wuthrich2012}, and Isham and Butterfield \cite{Butterfield:1998dd}, \cite{Butterfield:1999ah} to bring out a comprehensible philosophical characterization of quantum gravity.

	The term `loop' comes from the solution written for every line closed on itself on the proposed structure of quanta's interactions. John Archibald Wheeler was one of the pioneers  in constructing a representation of space which had a granular structure on a very small scale. Together with Bryce DeWitt they produced a mathematical formula known as Wheeler-DeWitt equation, ``an equation which should determine the probability of one or another curved space''. (\cite{Rovelli16}, loc. 1716) The starting point was spacetime of general relativity having ``loop-like states'' (\cite{Rovelli04}, p.13). Having a quantum approach to gravity on closed loops, which are threads of the Faraday lines of the quantum field,  constitutes a gravitational field which looks like a spiderweb. A solution could be written for every line closed on itself. Moreover, every line determining a solution of the Wheeler- DeWitt equation describes one of the threads of the spiderweb created by Faraday force lines of the quantum field which are the threads with which the space is woven. (cf. \cite{Rovelli16}, loc. 1757) 

	Space, then, is defined based on the nodes on this spiderweb, which is called a \textit{spin network}, and time, which already lost its fundamental status with special and general relativity, vanishes from the picture of the universe altogether.

	Loop quantum gravity combines the dynamic spacetime approach of general relativity with  quanta nature of gravity fields. Accordingly, space that bends and stretches are made up of very small particles which are called quanta of space.  If one had eyes capable zooming into the space and seeing magnetic fields and quanta, then, by observing the space, one would  first  witness the quantum field, and then would end up seeing quanta which are extremely small and granular.

	In the next subsection I give an account of space and time as they appear in the subject of loop quantum gravity.

	\subsection{Quantian Space and Time }

	Quantum mechanics describe the world we live in as a granular, fluctuating artifact. Planck constant determines the scale of this discrete differentiation. What does this say about space and time? It means that there is no continuum  for space and time. It signifies that space is made of quanta since it is a gravitational field. These quanta vibrate and exist only when there is an interaction.  
	By applying general equations of quantum mechanics to gravitational field, loop quantum theory describes the mathematical structure of  quanta form of space. As a result, we get a finite system of quanta which constitutes the space with with their minimum volume on Planck scale. 
	
	The way this mathematical form looks is a graph with nodes which describe the volume of  space.  The interactions by lines  are the units of space in the quantum sense. They are the quanta of space, which are quanta of gravity. They create space by appearing in that place. Hence,\textit{ place of the quanta of gravity} becomes the space as we have seen in relational theories of Aristotle and Leibniz. The place of quanta is characterized by  the interacting quanta next to each other. Therefore, this mathematical model suggests a pure relational view of space. A spin network is the resulting closed circuit which looks as the following:

	\begin{figure} [H]
		\centering
		\includegraphics[width=30mm]{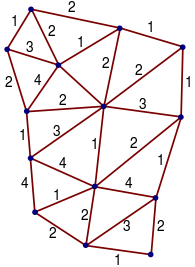}
		\caption{Simple spin network of the type used in loop quantum gravity \cite{Spin}}
	\end{figure}
	
	Since the structure is a closed circuit,  curvature can be measured and  the force of the gravitational field can be  determined for each loop.  (cf. (\cite{Rovelli04}, p.12), (\cite{Rovelli16}, loc. 1815)). Hence, the spin network describes the quantum state of the gravitational field.

	Spin networks evolve in a random manner, time counts the interaction of  nodes. Time becomes, as does space, only a consequence of these interactions. With general relativity time already lost its status as a fundamental concept. Time is defined as simultaneous events happening and that notion is localized. There has been no grandfather clock of universe since the postulation of special relativity. In quantum gravity, where Wheeler-DeWitt equation contains no time constant and  where the measurements are in extreme scale, now there is no more time. 
	
	The physical occurrences are then defined with quanta's initial and final states in addition to their possible trajectories between these two states. There is no concept of ``during'' but rather  possibilities of location points in the nodes. 
	
	Is there a history in this model? The transitions of spin networks between spin network states are called the history  which create a foam-like structure known as \textit{spin foam} (cf. \cite{Perez:2012wv}, p.29). Basically, a spin foam is shaped by the drifting spin network with the nodes splitting and merging as well. (cf. \cite{Rovelli16}, loc.2045)  This helps with computing the probability of any physical event. As Rovelli  summarizes  ``The microscopic swarming of quanta, which  generates space and time, underlies the calm appearance of macroscopic reality surrounding us. Every cubic centimeter of space, and every second that passes is the result of this dancing foam of extremely small quanta''. (\cite{Rovelli16}, loc.2081)
	
	Moreover, the interactions between quanta create the space, and time emerges as the possibility of these interactions. There is no past and future, the quanta only appear when there is interaction. This is similar to the relational view of space similar to the one Leibniz provided where space is not infinitely divisible.

	In the next section, I deliberate about the philosophical implications of space and time emerging from the loop quantum gravity. 
	
	\section{Kant and Quantum Gravity}
	
	As we have seen in Section 2, Kant argues that pure intuitions space and time are the\textit{ a priori} conditions of our understanding and experience. Time provides the continuous manifold for one to hold the thought and carry out reasoning; space provides the background for these thoughts to be exhibited. In the following, I discuss whether the changing notions of space and time due to quantum gravity make the Kantian characterization an invalid one.

	\subsection{Problem of Space and Time}
	
	As I have outlined in the previous section, space and time lose their fundamental status according to the model of loop quantum gravity. Spacetime in general relativity has already surprised us with bending space and localized time. Loop quantum gravity pushes the limits of our understanding since ``the idea that the universe is not in space and time arguably shocks our
	very idea of physical existence as profoundly as any scientific revolution ever did'' (\cite{Wuthrich2012}, pp.9-10) In this paper, I discuss the problem of space and time originating from loop quantum gravity with a broader philosophical perspective, and not ``the problem of time'', which stems from the incompatibility of  meaning of time in general relativity and quantum mechanics.  The approach of Rovelli I presented above already contains a solution to this specific problem of time by presenting the unification, that is, the loop quantum gravity, as a timeless theory. \cite{Anderson:2010xm} The perspective I present here deals with the issue of loosing the fundamental nature of space and time, whereby shifting our more intuitive continuous representations of them. My aim is to compare these emerging notions to the Kantian characterization of them. 
	
	There are two kinds of remarks from experts of quantum gravity about the Kantian characterization of space and time.  The first one, by Rovelli \cite{Rovelli16}, suggests that the Kantian approach to  space and time is mistaken (\cite{Rovelli16}, p. 169); and the second one, by Butterfield and Isham \cite{Butterfield:1999ah}, is  more sympathetic to  the ``idea that human understanding of reality must, as an a
	priori matter, involve certain notions of space and time'' (\cite{Butterfield:1998dd}, p.19)

	Kant asserts that time is not absolute, so does Leibniz \cite{Leibniz56},  Einstein \cite{Einstein06} and G\"{o}del\cite{Godel00}. The reason I mention G\"{o}del too is due to his interesting comment on Kantian notion of space and time. He conveys that ```the relativity theory is almost a verification of Kant’s doctrines'' (\cite{Godel95a}, p. 230). 
	
	G\"{o}del acknowledges that  ``the trend of modern physics is in one respect opposed to Kantian philosophy.'' However, he states that, ``it should not be overlooked that the very refutation of Kant's assertion concerning the impossibility for theoretical science of stepping outside the limits of our natural conception of the world has furnished in several points a most striking confirmation of his main doctrine concerning this natural world picture, namely, its largely subjectivistic character, even as to those concepts which seem to constitute the very back bone of reality.''  (\cite{Godel95a}, p. 246) G\"{o}del is well aware that Kantian notions of space and time have only a phenomenological reality, however, they still provide the objective grounds of our reality, which is the most important notion about Kant's phenomenology.  
	
	Furthermore, he specifically points out that ``I do not think the question whether, in accordance with Kant's view we have an \textit{innate} (and therefore a priori) intuition of Euclidean space (i.e., whether we would develop the same intuition also in a strongly non-Euclidean world) has yet been decided; nor the related question whether we are able (in our world) to learn to imagine a non-Euclidean space. [For we can imagine this also in terms of regular changes of size and shape due to motion in a Euclidean space.]... But whatever the answers to these questions may be, there can certainly never result from them any incompatibility between Kant and relativity theory, but at most between Kant and psychology (or phenomenology) of sense perception.'' (\cite{Godel95a}, pp.243-244)  
	
	Cognitive researchers provide an answer to the first part of G\"{o}del's inquiries. The new analysis and experiments indicate that we have an innate tendency to conceptualize in Euclidean space. Izard et al. claim that even if human perception of space ``violate Euclidean principles in several ways (\cite{Hatfield03}, \cite{Koen10}) (for example, by imposing a curvature to the space (\cite{Indow91}) or failing to unify different scales (\cite{Lune47})), the way that we conceive space is not necessarily constrained by our perception.'' (\cite{Izard11}, p.1) They argue, ``following Kant's proposal, that the axioms of Euclidean geometry may constitute the most intuitive conceptualization of space not only in adults educated in the tradition of Euclidean geometry (cf. \cite{Asmu06}) but also in cultures where this tradition is absent.''  (\cite{Izard11}, p. 1) They point out that: ``Euclidean geometry, inasmuch as it concerns basic objects such as points and lines on the plane, is a cross-cultural universal that results from inherent properties of the human mind as it develops in its natural environment'' (\cite{Izard11}, p. 4) and ``the basic principles of Euclidean geometry are reflected in intuitions of space that develop progressively throughout childhood, but still appear universal''.(\cite{Izard2011319}, p. 330)  These results from cognitive science not only provide an answer to G\"{o}del's inquiry but also confirm the Kantian thesis that Euclidean space, as it inhibits our cognition due to the formation of human mind, provides an \textit{a priori} condition for our experience. Hence, there is backing up from cognitive research that Kantian thesis of space is not necessarily overthrown by discoveries of improved logic systems or new spaces that are not Euclidean.

	G\"{o}del's final remark on the comparison between Kantian characterization of physical notions and relativity theory indicates  a similar consequence for space and time as they appear (or not) in loop quantum gravity. In other words, Kant's system is not influenced by discovery of discrete spaces, providing the subjective nature of time or postulating that there is no time. Kantian characterization of space and time explains what we can know phenomenologically with certainty. It is not an attempt to know or describe a space and time outside of human cognition.
	
	Kant regards his approach  as a Copernican revolution in philosophy, only in this case space and time were pulled back to the human mind. (cf. \cite{Kant98}, Bxvii) To Kant, physical reality is something we construct in our minds, however, since we have \textit{a priori} the condition of Euclidean geometry, this subjective approach delivers an objective reality in the case of pure and \textit{a priori} representations such as space, time and mathematical constructions.   Hence, G\"{o}del is right about pointing out that unless the characterization of space and time stems from phenomenology or cognitive science, one cannot compare it with Kantian notions of them. Then an interesting question for loop quantum gravity is whether we can have a phenomenological approach to the notions emerging from it.

A comparable theory of loop quantum gravity to the Kantian characterization needs to account for space and time as the basis of a human faculty. I believe, we have a candidate, which is called ``intuition''.   

	Carlo Rovelli proposes that maybe we should not question what to  change about quantum mechanics but rather question ``what is limited about our intuition that makes it seem so strange to us'' (\cite{Rovelli16}, loc. 1568) He proposes that because of our limited imagination the quantum mechanics seem obscure to us. 
	
	I think,  we should  reframe this question by replacing intuition with reasoning.   Because these results might not be strange to   our intuition  but to our reasoning, which is based on the information we learn and the education we get. If we become an outside observer of our thoughts and emotional states we can grasp the notion of time lapse and space disappearing, however, reasoning is always continuous and needs space and time in Newtonian sense.
	
First, let me point out that, in this section,  I  use intuition in a broader sense of \textit{appearing immediately in consciousness} following Gary Klein \cite{Klein98}
	 and not in  Kantian sense that refers to singular constructions. Kant uses intuition as a technical term for his transcendental philosophy and it is almost never accounted as a propositional attitude. Hence, we need to depart from the Kantian schema when talking about intuition as a propositional attitude or a human faculty. We can describe intuition further by characterizing it as  ``a form of knowledge that is accessible to explicit report, although its justification is not.''(\cite{Izard2011319}, p. 331)

	   Kant uses Newtonian framework to describe space similar to a mental sheet for our singular representations to be exhibited and time to hold the sequence of these representations so that one can form a thought using them as bases. A theory of loop quantum gravity then  would have the thought precede  causality, time and space. 
	 
	  In loop quantum gravity the trajectory a quantum particle follows is calculated by summing over all the possible trajectories of probability cloud as I talked about in Section 3. Recall that I also pointed out that in loop quantum gravity there is  no continuous process, no ``during''. There are  initial and final states but the path  quantum particle follows is not deterministic. Intuitive conclusions seem to follow a similar pattern, by presenting a response and not being able to explain the justification. It seems that the conclusion does not follow a certain path but rather intuition scans  possible candidates to the problem and picks one of the paths leading to the answer. There is similarly no ``during'' in this process which indicates that  intuition does not use time. Intuition also does not locate previous thoughts and tries to arrange them to arrive at a conclusion as well which makes the space secondary nature. Hence, similar to quantum particles, with the intuitive process, thought appears first, preceding space and time. I believe this can be explained due to  having two excellent theories defining universe, one with wave functions one with quanta.  It is not then a far possibility for us human beings have two aspects that define our consciousness. A matter one and a vibration one, the one conforming the laws of Newtonian physics and the other conforming the laws of general relativity and quantum physics. Former describing how we think and reason, the latter providing us clues about how intuition functions.

 \section{Conclusion}
	 
	Quantum gravity tells us that space and time are not fundamental concepts. Gravity field with quanta creates space according to loop quantum gravity. Existence of quanta is only possible with interactions and time emerges through this process. Although this seems to contradict what Kant has been describing, as I have discussed above, in my opinion, it does not. Because this approach still does not affect the way we perceive and experience outer world and how space and time are the\textit{ a priori} condition of our understanding.

	Reasoning and logic need Kantian space and time in order to proceed to conclusions and exhibit intuitions in Kantian sense. When we think about  quanta, or ponder upon its possibility of discontinuity we exhibit these notions in time, albeit an internal one, and we use a mental sheet to visualize these concepts. These are the Kantian pure intuitions space and time we use. Although Kant based his theory on Newtonian physics,  he never claimed that there is an absolute space as Newton did. His aim was to improve Newton's physics by using Leibnizian- Wolfian framework of space and time, a relational theory as I described in Section 2. Hence, unless we develop a cognitive theory of quantum gravity suggesting  how one can reason in \textit{no time} and exhibit representations in \textit{a space that comes after  interactions} we do not have a comparable  characterization with Kant's theory. Therefore, as I pointed out, when physicists of quantum theory suggest that Kantian space and time are partially or fully overthrown  due to the results of loop quantum gravity, in my opinion, this comparison is not valid. Neither loop quantum gravity gives a constraint on Kantian characterization of space and time nor Kant's notions of them  to the quantum gravity or any  theory of the physics at the very least.

	\bibliographystyle{plain}
	\bibliography{extendibility}

\begin{thebibliography}{10}

\bibitem{Anderson:2010xm}
Edward Anderson.
\newblock {The Problem of Time in Quantum Gravity}.
\newblock 2010.

\bibitem{Aristo83}
Aristotle.
\newblock {\em Physics: Books III and IV}.
\newblock Oxford University Press, USA, 1983.

\bibitem{Asmu06}
Jennifer~A. Asmuth and Lance~J. Rips.
\newblock Conceptual change in non-{E}uclidean mathematics.
\newblock {\em Annual Conference of the Cognitive Science Society}, pages
  30--35, 2006.

\bibitem{Butterfield:1998dd}
J.~Butterfield and C.~J. Isham.
\newblock {On the emergence of time in quantum gravity}.
\newblock 1998.

\bibitem{Butterfield:1999ah}
J.~Butterfield and C.~J. Isham.
\newblock {Space-time and the philosophical challenge of quantum gravity}.
\newblock In {\em In *Callender, C. (ed.) et al.: Physics meets philosophy at
  the Planck scale* 33-89}. 1999.

\bibitem{Einstein06}
Albert Einstein.
\newblock Zur elektrodynamik bewegter k\"{o}rper.
\newblock {\em Annalen der Physik}, 17:891--921, June 1905.

\bibitem{Friedman94}
Michael Friedman.
\newblock {\em Kant and the Exact Sciences}.
\newblock Harvard University Press, London, 1994.

\bibitem{Godel95a}
Kurt G\"{o}del.
\newblock Some observations about the relationship between theory of relativity
  and kantian philosophy.
\newblock In Solomon Feferman, editor, {\em Kurt G\"{o}del: Colected
  Works,Volume III, Unpublished essays and lectures}, pages 230--247. Oxford
  University Press, 1995.

\bibitem{Godel00}
Kurt G\"{o}del.
\newblock An example of new type of cosmological solution of einstein's field
  equations of gravitation.
\newblock {\em General Relatity and Gravitation}, 32(7):1409--1417, 2000.

\bibitem{Hatfield03}
Gary Hatfield.
\newblock Representation and constraints: The inverse problem and the structure
  of visual space.
\newblock {\em Acta Psychol}, 114:355--378, 2003.

\bibitem{Indow91}
Tarow Indow.
\newblock A critical review of {L}uneburg's model with regard to global
  structure of visual space.
\newblock {\em Psychol Rev}, 98:430--453, 1991.

\bibitem{Izard11}
V.~Izard, P.~Pica, E.S. Spelke, and S.~Dehaene.
\newblock Flexible intuitions of {E}uclidean geometry in an {A}mazonian
  indigene group.
\newblock {\em PNAS}, 108(24):9782--9787, 2011.

\bibitem{Izard2011319}
V\'{e}ronique Izard, Pierre Pica, Stanislas Dehaene, Danielle Hinchey, and
  Elizabeth Spelke.
\newblock Geometry as a universal mental construction.
\newblock In Stanislas Dehaene and Elizabeth~M. Brannon, editors, {\em Space,
  Time and Number in the Brain}, pages 319 -- 332. Academic Press, San Diego,
  2011.

\bibitem{Kant1902a}
Immanuel Kant.
\newblock {\em AKK. Kants gesammelte Schriften}.
\newblock Berlin, 1902-.

\bibitem{Kant98}
Immanuel Kant.
\newblock {\em Critique of Pure Reason}.
\newblock Cambridge University Press, Cambridge, 1998.

\bibitem{Klein98}
Gary Klein.
\newblock {\em Sources of Power}.
\newblock Cambridge University Press, Cambridge, 1998.

\bibitem{Koen10}
Jan~J. Koenderink~et al.
\newblock Does monocular visual space contain planes?
\newblock {\em Acta Psychol}, 134:40--47, 2010.

\bibitem{Leibniz56}
Gottfried~Wilhelm Leibniz.
\newblock {\em Philosophical Papers and Letters}.
\newblock University of Chicago Press, Chicago, 1956.

\bibitem{Lune47}
Rudolf~Karl Luneburg.
\newblock {\em Mathematical Analysis of Binocular Vision}.
\newblock Princeton University Press, Princeton, 2010.

\bibitem{Martinetti:2012jh}
Pierre Martinetti.
\newblock {Emergence of time in quantum gravity: Is time necessarily flowing?}
\newblock 2012.

\bibitem{Newton66}
Isaac Newton.
\newblock {\em Sir Isaac Newton's Mathematical Principles of Natural
  Philosophy}.
\newblock University of California Print, Berkeley, 1966.

\bibitem{Perez:2012wv}
Alejandro Perez.
\newblock {The Spin Foam Approach to Quantum Gravity}.
\newblock {\em Living Rev. Rel.}, 16:3, 2013.

\bibitem{Plato2000}
Plato.
\newblock {\em Timaeus}.
\newblock Hackett Publishing, Indianapolis and Cambridge, Mass, 2000.

\bibitem{Rovelli04}
Carlo Rovelli.
\newblock {\em Quantum Gravity}.
\newblock Cambridge University Press, Cambridge, 2004.

\bibitem{Rovelli16}
Carlo Rovelli.
\newblock {\em Reality is Not What It Seems}.
\newblock Allen Lane, Great Britain, 2016.

\bibitem{Space3}
Wikipedia.
\newblock {Euclidean Space}.
\newblock \url { http://en.wikipedia.org/wiki/EuclideanSpace}, 2015.

\bibitem{Space6}
Wikipedia.
\newblock {Non Euclidean Geometry}.
\newblock \url {http://en.wikipedia.org/wiki/NonEuclideanGeometry/}, 2015.

\bibitem{Space4}
Wikipedia.
\newblock {Spacetime}.
\newblock \url { http://en.wikipedia.org/wiki/Spacetime}, 2015.

\bibitem{Spin}
Wikipedia.
\newblock {Spin Network}.
\newblock \url
  {https://en.wikipedia.org/wiki/Spin_network#/media/File:Spin_network.svg},
  2016.

\bibitem{Wolff99}
Christian Wolff and J.~E. Hoffmann.
\newblock {\em Anfangsgr\"{u}nde Aller Mathematischen Wissenschaften (Latin:
  Elementa matheseos universa 1713-1715}.
\newblock G. Olms, Hildesheim, 1999.

\bibitem{Wuthrich2012}
Christian Wuthrich.
\newblock {In search of lost spacetime: philosophical issues arising in quantum
  gravity}.
\newblock {\em In Soazig}, LeBihan:Pr\'{e}cis de philosophie de la physique,
  Paris: Vuibert (2013), 2013.

\end{thebibliography}
\end{document}